\documentstyle[editedvolume,epsf]{crckapb}

\begin{opening}
\title{HIGHLIGHTS: OPTICAL/NIR SPECTROSCOPY \protect\\
       OF ULTRALUMINOUS INFRARED GALAXIES}
\author{S. VEILLEUX}
\institute{University of Maryland\\
           College Park, MD 20742 USA}

\end{opening}

\runningtitle{OPTICAL/NIR SPECTROSCOPY OF ULIGS}

\def\eps@scaling{.95}
\def\plotone#1{\centering \leavevmode
\epsfxsize=\eps@scaling\columnwidth \epsfbox{#1}}
\def\lesssim{\mathrel{\hbox{\rlap{\hbox{\lower4pt\hbox{$\sim$}}}\hbox{$<$}}}}
\def\gtrsim{\mathrel{\hbox{\rlap{\hbox{\lower4pt\hbox{$\sim$}}}\hbox{$>$}}}}

\begin{document}

\noindent{\bf Abstract.}
This paper reviews the results from recent optical and near-infrared
spectroscopic studies of ultraluminous infrared galaxies.

\section{Introduction}

Over the last decade, several spectroscopic studies have attempted to
determine the nature of the dominant energy source in ultraluminous
infrared galaxies (ULIGs).  Optical surveys of unbiased samples of
luminous {\it IRAS} galaxies (e.g. Elston, Cornell, \& Lebofsky 1985;
Leech et al. 1989; Allen et al. 1991; Ashby, Houck, \& Matthews 1995;
Wu et al. 1998) generally have found that $\gtrsim$ 80\% of
high-luminosity infrared galaxies (LIGs; $L_{\rm ir} > 10^{11}\
L_\odot$) present H II region-like optical spectra, and therefore
appear to be powered by hot stars rather than an active galactic
nucleus (AGN).  However, the great majority of the infrared galaxies
in these samples are in the luminosity range $L_{\rm ir} =
10^{11}-10^{12}\ L_\odot$, with only a few having $L_{\rm ir} >
10^{12}\ L_\odot$. This distinction is important as there is growing
evidence that the fraction of AGN among LIGs increases with increasing
$L_{\rm ir}$ (e.g., Sanders et al. 1988; Armus, Heckman, \& Miley
1989; Veilleux et al. 1995; Sanders \& Mirabel 1996; cf.~\S 2).

Recent progress in infrared detector technology provides another
approach to constrain the energy source in ULIGs. Near-infrared
spectroscopy has the potential to more deeply probe the cores of
ULIGs: for example, the extinction coefficient in the K-band is nearly
10 times smaller than at optical wavelengths.  This technique has
proven very useful in the study of highly reddened broad-line regions
(BLRs) in intermediate Seyferts (1.8's and 1.9's; Goodrich 1990; Rix
et al. 1990) and also has had success finding obscured BLRs in some
optically classified Seyfert~2 and radio galaxies (e.g., Blanco, Ward,
\& Wright 1990; Goodrich, Veilleux, \& Hill 1994; Ruiz, Rieke, \&
Schmidt 1994; Hill, Goodrich, DePoy 1996; Veilleux, Goodrich, \& Hill
1997a) and in a few ULIGs (e.g., DePoy et al. 1987; Hines 1991; Nakajima,
et al. 1991a,b; see Goldader et al. 1995, however).

The recent publication of the 1-Jy sample of ULIGs (Kim \& Sanders
1998) offers a unique opportunity to verify these
optical/near-infrared spectroscopic results.  The 1-Jy survey provides
a complete list of the brightest ULIGs with $f_{60}$ $>$ 1 Jy which is
not biased toward `warm' quasar-like objects with large
$f_{25}/f_{60}$ ratios.  This sample contains 118 objects with $z$ =
0.02 -- 0.27 and log~[L$_{\rm ir}$/L$_\odot$] = 12.00 -- 12.90.  The
infrared luminosities of these objects therefore truly overlap with
the bolometric luminosities of optical quasars.  Other surveys have
discovered objects of comparable luminosity at fainter flux levels as
well as a few `hyperluminous' objects at higher L$_{\rm ir}$. However,
the 1-Jy sample contains the brightest objects at a given luminosity,
hence the best candidates for follow-up studies.  The results from our
optical spectroscopic survey of this sample are presented in Kim,
Veilleux, \& Sanders (1998; KVS) and Veilleux, Kim, \& Sanders (1999a;
VKS) and are summarized in \S 2. The near-infrared survey was just
recently completed (Veilleux, Sanders, \& Kim 1999b; VSK); the results
are summarized in \S 3.  The last section of the present paper (\S 4)
combines the results from the optical and near-infrared surveys to
derive the frequency of AGN/quasars among ULIGs and evaluate the
importance of AGN/quasars in powering the large bolometric
luminosities of ULIGs.

\section{Optical Spectroscopy of the 1-Jy Sample of ULIGs}

KVS and VKS discuss the optical spectroscopic properties of the 1-Jy
sample of 118 ULIGs.  These spectra are combined with those of
Veilleux et al. (1995) to look for systematic trends with infrared
luminosity among LIGs with $L_{\rm ir} \approx 10^{10.5}-10^{13}\
L_\odot$.  As found in Veilleux et al. (1995), the fraction of Seyfert
galaxies among LIGs increases with infrared luminosity.  For $L_{\rm
ir} > 10^{12.3}\ L_\odot$, about 48\% of the ULIGs (15/31 objects) are
classified as Seyfert galaxies.  A summary of the spectral
classification as a function of the infrared luminosity is presented
in Figure 1.

\begin{figure}[htbp]
\includegraphics{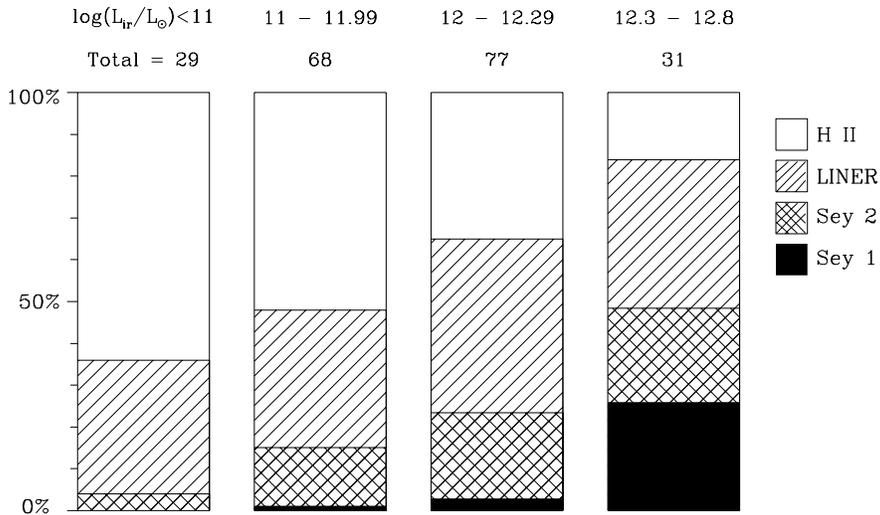}
\centerline{}
\vskip 3.5in
\caption{Optical spectral classification as a function of infrared luminosity
for the objects in the 1-Jy sample and the luminous infrared galaxies
in the Bright Galaxy Survey (Veilleux et al. 1995, 1999a; Kim et al. 1998).}
\end{figure}

Many of the spectroscopic properties of the Seyfert galaxies point to
the existence of an AGN which is not present or visible in LINER or
H~II ULIGs.  About 30\% (10/33) of the Seyfert galaxies in the 1-Jy
sample are of type 1, presenting broad Balmer lines and strong Fe~II
emission similar to what is observed in optically selected quasars.
Seyfert ULIGs (especially those of type 1) have weaker H$\beta$ and
Mg~Ib stellar absorption features, bluer continuum colors, larger
H$\alpha$ luminosities and equivalent widths, smaller
infrared-to-H$\alpha$ luminosity ratios, and warmer {\em IRAS} ${\em
f}_{25}/{\em f}_{60}$ colors than LINER or H~II ULIGs.  The [O~III]
$\lambda$5007 line widths in the nuclei of the Seyfert galaxies are
also significantly broader on average than those measured in the H~II
and LINER ULIGs.

The weak H$\beta$ and Mg~Ib features in ULIGs optically classified as
H~II galaxies or LINERs suggests the presence of a young ($\lesssim$
few $\times$ 10$^7$ yrs) stellar population comprising $\sim$ 10\% of
the total galaxy mass in these objects (Bica, Alloin, \& Schmidt
1990).  While this starburst is the likely source of ionization among
H~II galaxies and some LINERs, long-slit information from Veilleux et
al. (1995) and KVS indicates that the LINER-like
emission may also be produced through shocks caused by the interaction
of starburst-driven outflows with the ambient material.  The weaker
stellar absorption features, bluer observed continuum colors, and
larger H$\alpha$ emission equivalent widths among ULIGs indicate that
the starburst becomes increasingly important with increasing infrared
luminosity in both H~II galaxies and LINERs.


\section{Near-Infrared Spectroscopy of the 1-Jy Sample of ULIGs}

The latest results from a sensitive near-infrared search for obscured
BLRs in ULIGs from the 1-Jy sample are presented in VSK. 
The results from this survey were combined with those
obtained by Veilleux, Sanders, \& Kim (1997b) to produce a
near-infrared spectroscopic database on sixty-four ULIGs.  Prior to
selecting these objects, the $\sim$ 10\% of the 118 galaxies that
already were known optically to show direct signs of quasar activity,
i.e. optically classified as Seyfert~1, were excluded.
VSK find that all of the galaxies with strong
evidence for a hidden BLR at near-infrared wavelengths (Pa$\alpha$
and/or Pa$\beta$) present an optical Seyfert~2 spectrum.  Overall,
50\% (and perhaps up to 70\%) of the optical Seyfert~2 galaxies in the
combined sample present either a BLR or strong [Si~VI] emission.  All
ten `warm' ($f_{25}/f_{60} > 0.2$) optically classified Seyfert 2
galaxies in the sample show either obscured BLRs or [Si~VI] emission
at near-infrared wavelengths. None of these objects have deficient
Pa$\alpha$-to-infrared luminosity ratios.  These results strongly 
suggest that the screen of dust obscuring the cores of `warm' Seyfert
2 ULIGs is optically thin at 2~$\mu$m.  In contrast, none of the 41
optically classified LINERs and H~II galaxies in the sample shows any
obvious signs of an energetically important AGN at near-infrared
wavelengths.  The LINERs and H~II galaxies in the sample span a wide
range of {\it IRAS} colors and Pa$\alpha$-to-IR luminosity ratios. The
apparent lack of AGN activity in these objects is therefore unlikely
to be due solely to dust obscuration.

\section{Discussion and Conclusions}

The optical and near-infrared data taken together, suggest
that the total fraction of objects in the 1-Jy sample with signs of a
bonafide AGN is at least $\sim$ 20 -- 25\%.  This fraction reaches 35
-- 50\% for objects with $L_{\rm ir} > 10^{12.3}\ L_\odot$.  These
percentages are lower limits because the near-infrared method often
fails to detect AGN activity ([Si~VI] emission) in known optically
selected Seyfert 2 galaxies (Marconi et al. 1994).

The presence of AGN activity in ULIGs does not necessarily imply
that the AGN is the dominant energy source in these objects. A more
detailed look at the AGN in these ULIGs is needed to answer this
question. Following Veilleux et al. (1997b), we have plotted in Figure
2 the dereddened {\em broad-line} H$\beta$ luminosities of the optical
and obscured BLRs in ULIGs and optically identified QSOs as a function
of their bolometric luminosities. The methods and assumptions which
were used to create this figure are briefly described in the caption
(cf. Veilleux et al. 1997b, VKS, and VSK for a more detailed
discussion).  The typical uncertainties on the data points of Figure 2
are of order $\pm$ 30\%.  This figure brings support to the idea first
suggested by Veilleux et al. (1997b) that most ($\sim$ 80\%) of the
ULIGs with optical or near-infrared BLRs in the 1-Jy sample are
powered predominantly by the quasar rather than by a powerful
starburst.  The AGN/quasar is therefore the dominant energy source in
at least 15 -- 25\% of all ULIGs in the 1-Jy sample. This fraction is
closer to 30 -- 50\% among ULIGs with $L_{\rm ir} > 10^{12.3}\
L_\odot$. ULIGs with powerful AGN/quasar but with highly obscured BLRs
would increase these percentages.  An object-by-object comparison of
our optical/near-infrared results with those obtained with ISO
indicates a very good agreement between these two sets of data
(cf.~Genzel et al. 1998; VSK; Lutz, Veilleux, \&
Genzel 1999).


\acknowledgements{ The ground-based study discussed in this paper was
done in collaboration with Drs. D. B. Sanders and D.-C. Kim.  The
author gratefully acknowledges the financial support of NASA through
LTSA grant number NAG~56547.

\begin{figure}[htbp]
\plotone{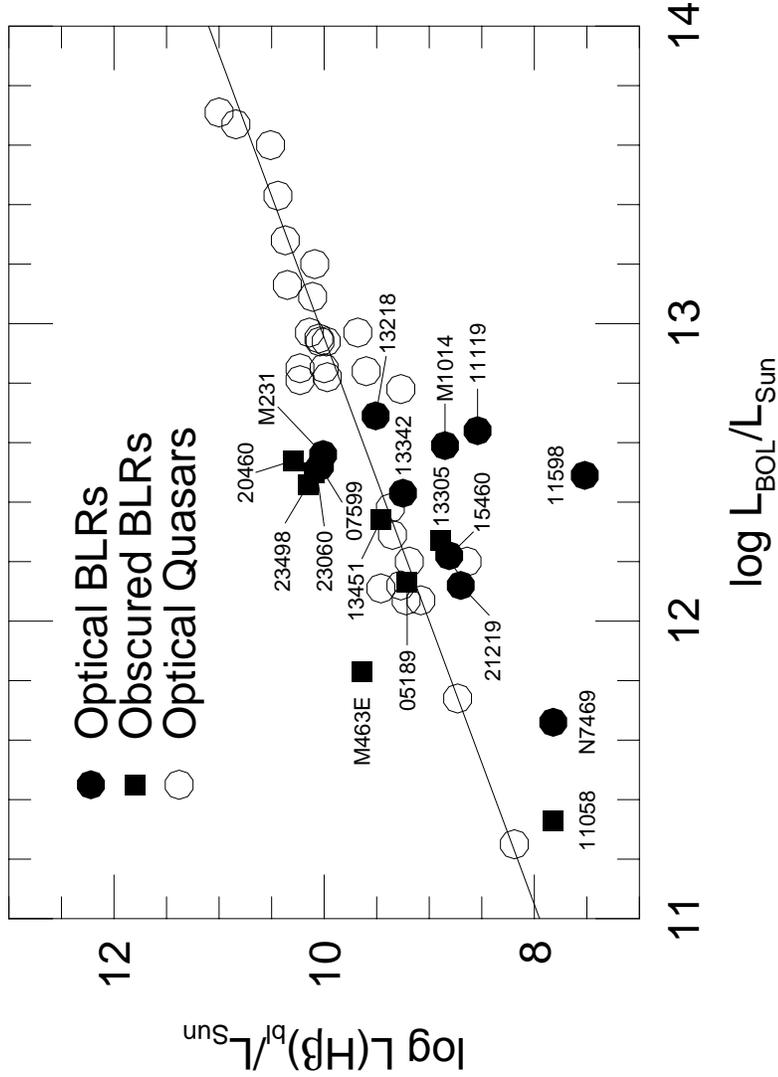}
\caption{Comparison of the H$\beta$ luminosities of optical and
obscured BLRs in ULIGs and optically identified QSOs as a function of
their bolometric luminosities. $L_{\rm bol}$ for the QSOs was
determined using the bolometric correction factor 11.8 (i.e. $L_{\rm
bol} = 11.8 \nu_B L_\nu(B)$: Elvis et al. 1994; Sanders \& Mirabel
1996), except for those few sources that were detected by {\it IRAS},
in which case $L_{\rm bol}$ was taken from Sanders et al. (1989).  For
the ULIGs $L_{\rm bol}$ was taken to be $1.15 \times L_{\rm ir}$ (Kim
\& Sanders 1998).  The H$\beta$ data for the optically selected QSOs
are from Yee (1980) corrected for $H_{\rm o} =
75$~km~s$^{-1}$~Mpc$^{-1}$ and $q_{\rm o} = 0$.  The H$\beta$
luminosities of the ULIGs with obscured BLRs were calculated from the
measured broad Pa$\alpha$ fluxes assuming Case B recombination (except
for Mrk~463E where the broad Pa$\beta$ flux was used).  The reddening
correction was carried out using the color excesses derived from the
infrared line ratios. Most of the ULIGs fall close to the quasar
relation (solid line). }
\end{figure}

\end{document}